# A case for multiple and parallel RRAMs as synaptic model for training SNNs

Aditya Shukla
*Dept. of Electrical Engineering*
*Indian Institute of Technology Bombay*
Mumbai, India
adityashukla@ee.iitb.ac.in

Sidharth Prasad
*Dept. of Electrical Engineering*
*Indian Institute of Technology Bombay*
Mumbai, India
prasad.sidharth@columbia.edu

Sandip Lashkare
*Dept. of Electrical Engineering*
*Indian Institute of Technology Bombay*
Mumbai, India
sandipl@ee.iitb.ac.in

Udayan Ganguly
*Dept. of Electrical Engineering*
*Indian Institute of Technology Bombay*
Mumbai, India
udayan@ee.iitb.ac.in

*Abstract*— To enable a dense integration of model synapses in a spiking neural networks hardware, various nano-scale devices are being considered. Such devices, besides exhibiting spike-time-dependent plasticity (STDP), need to be highly scalable, have a large endurance and require low energy for transitioning between states. In this work, we first introduce and empirically determine two new specifications for a synapse in SNNs: number of conductance levels per synapse and maximum learning-rate. To the best of our knowledge, there are no RRAMs that meet the latter specification. As a solution, we propose the use of multiple RRAMs in parallel within a synapse. While synaptic reading, all PCMO-RRAMs are simultaneously read and for each synaptic conductance-change event, the mechanism for conductance STDP is initiated on only one RRAM, randomly picked from the set. Second, to validate our solution, we experimentally demonstrate STDP of conductance of a PCMO-RRAM and then show that due to a large learning-rate, a single PCMO-RRAM fails to model a synapse in the training of an SNN. As anticipated, network training improved as more PCMO-RRAMs are added to the synapse. Fourth, we discuss the circuit-requirements for implementing such a scheme, to conclude that the requirements are within bounds. Thus, our work presents specifications for synaptic devices in trainable SNNs, indicates the shortcomings of state-of-art synaptic contenders, and provides a solution to extrinsically meet the specifications and discusses the peripheral circuitry that implements the solution.

## I. INTRODUCTION

While decoding a brain's functioning, mimicking biology using models of brain has been a significant approach taken by neuro-scientific community. Also, to perform cognitive computing tasks while maintaining energy efficiency of a human brain has been the grand challenge in the post-scaling and data-driven era. Executing these tasks using spiking (or, third generation) neural networks in a bio-inspired hardware that is functionally and structurally similar to a brain will greatly help in achieving these goals. Such bio-mimicry has led to adoption of a bottom-up approach that distributes and integrates the processing and memory – electronic neurons as computing units and electronic synapses as memory units connected as required in network. Though various VLSI-amenable circuits have been proposed that can mimic a synapse [1], [2], to integrate as any many as $10^{14}$ synapses within a volume of a human brain requires the model synapse to have dimensions in the order of the thickness of a synaptic cleft (~10nm) [3]–[5]. This has led to an appreciable research and development of plethora of novel nano-scaled devices that can faithfully mimic a synapse [4], [6]–[17]. Of all the novel material based nano-scaled devices for modeling synapses in hardware, memristors/RRAMs (Resistive Random-Access-Memory) have gained remarkable research interest. These are non-volatile resistive memory-elements whose state/resistance can altered by applying sufficiently strong voltage-pulses and are strong candidates as weights in electronic in-situ trainable neural networks.

Spike-time dependent plasticity (STDP) rule, a type of Hebbian 'local' learning-rule, is considered to be the essential property of any SNN based synapse that their models must exhibit. By changing the strength of voltage-pulses being applied as a function of time-since-last-spike, STDP rule has been demonstrated on various RRAMs [4], [10]–[17]. But, for an extensive and large-scale utilization of these devices in cross-bars as a synaptic array, it needs to (1) be highly scalable, (2) have excellent endurance, (3) have low-energy switch-ability and (4) be compatible with CMOS [4], [18], [19]. However, other important but relatively unexplored requisites from the synaptic RRAM that are discussed in this work include: (1) analog range of conductance or ample number of memory states/bits and (2) a *low* value of maximum STDP based weight-change (mathematically, low $\max_{\Delta t}[|\Delta G(\Delta t)|/G]$) at each weight-change event, or learning-rate. Requisite (1) is based on the fact that in nature, most STDP rules observed in biology are analog [20], [21] and so are the data-sets (e.g. data-sets of images in color/grey-scale, Fisher's Iris, Wisconsin's breast-cancer, chemical assays like wine [22]–[24].) With only additional costs can either be synthetically transformed into the binary domain, thus often necessitating analog or multi-level synapse. Requisite (2) comes from the fact that all STDP learning-rules have a point of maximum weight-change (at point(s) of highest/least time or rate-correlation; [20], [21]), and this value should be kept small for a stable weight-evolution in a network while training.

In this paper, we first empirically show that for software-equivalent training, (1) the learning-rate (maximum $\Delta G/G$ at each weight-change event) must be less than 2% and (2) a resolution of at least 256 levels (8 bits) per synapse are needed. Second, we show that STDP demonstrations on RRAMs up-to-date depict a learning-rate of 20%-400%, thus, these devices do not meet our specifications. To tackle this problem we propose the use of multiple ($n$) and parallel RRAMs in a synapse. Within such a synapse, reading requires all RRAMs but the conductance-change, as dictated by the STDP rule, is brought about in only one randomly picked RRAM from the set of $n$ RRAMs of the synapse. This way, the learning-rate is lowered from $\Delta G/G$ to $\Delta G/nG$, enabling a software-equivalent learning. Second, we validate our proposal using an interpolation-model of STDP measurements of a standalone PrCaMnO-RRAM (accompanying all non-idealities) for training an SNN with multiple and parallel RRAMs in all its synapse. Next, the learning performance with $n$ is evaluated to show that $n = 4$ produces excellent peak learning performance but with significant fluctuations from epoch to epoch, while $n = 64$ is necessary for software-equivalent learning. In comparison, 256 binary synapses are needed for equivalent programming (4x improvement). Fourth, architectural consideration of circuit implementation is then discussed and a simple circuit to implement the random programming scheme is presented. Thus, our work presents the specification for synaptic devices for analog datasets, demonstrates that challenge for synaptic candidates in literature, and presents an architectural solution to enable learning and provides a circuit implementation.

This work is organized as follows: in section 2, we an overview of the STDP rule, followed by the basis for our claim of 2% learning-rate as a necessary condition for acceptable SNN training performance. In section 3 we report the procedure for STDP demonstration of the PCMO-RRAM and the results. In continuation, using the STDP-data we show that the this device fails as synapse. In section 4, we validate our proposal of using multiple and parallel RRAMs in a synapse. Lastly, in section 5, we discuss the hardware requirements and other consideration for adopting the proposed approach.

## II. IDEAL LEARNING-RATE FOR SNNS

### A. STDP-Overview

Spike-time dependent plasticity (STDP) is the most common learning-rule used in Spiking Neural Networks (SNNs). It gives a relation between $\Delta t$ (the time gap between the pre-synaptic and the post-synaptic spikes) and $\Delta G$, the weight change of the synapses, as illustrated in Figure 1.

For its illustration, we use an exponential weight-update rule comprising of exponential-$\Delta t$ dependence term, saturation factor $S_\pm(G)$ and scaling factors ($A_\pm$), as given in Eq. 2.

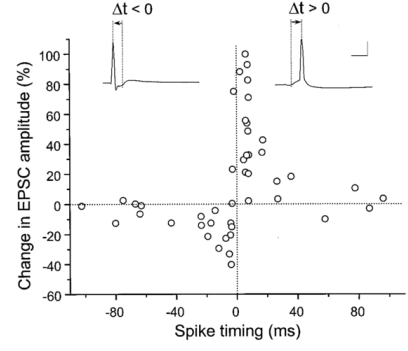

Figure 1: Biological STDP rule (after [20])

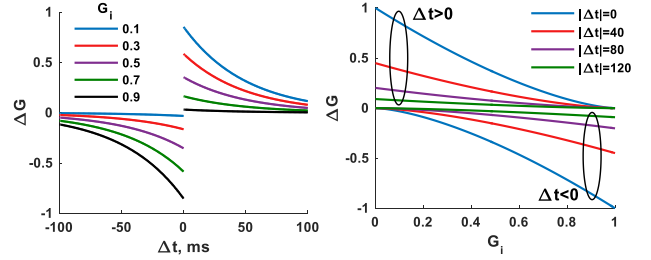

Figure 2: STDP rule of Eq. 1 with $S(G)$ in Eq. 3

$$\Delta G(\Delta t, G) = \begin{cases} A_+ \cdot S_+(G) \cdot e^{\frac{-\Delta t}{\tau_+}}, & \Delta t < 0 \\ A_- \cdot S_-(G) \cdot e^{\frac{\Delta t}{\tau_-}}, & \Delta t < 0 \end{cases} \quad (1)$$

Several other time-dependences of weight-change are possible [21]. $S_\pm(G)$ model the biological saturation-effects [20] and ensures that $G$ doesn't increase/decrease indefinitely. In Eq. 1, $A_+$ is positive while $A_-$ is negative (to strengthen causality and weaken anti-causality between spikes). As an example, we have plotted in Figure 2: (1) $\Delta G$ a function of $\Delta t$ for various $G_i$ (or $G$) and (2) $\Delta G$ as a function of $G_i$ (or $G$) for various $\Delta t$ with $\tau_\pm = 50$ms, $G_{max} = 1$, $G_{min} = 0$ and $A_\pm = 1$ and:

$$S_+(G) = (1-G)^{1.5} \ \& \ S_-(G) = G^{1.5} \quad (2)$$

### B. Ideal learning-rate for training

Since $A_\pm$ in Eq. 1 (or any STDP equation for that matter) set the maximum conductance-change, they determine any network's learning-rate. While training an artificial neural network, the learning-rate needs to be carefully chosen [25]. Since no work exists that studies learning-rate for spiking neural networks trained using STDP, we empirically determine the learning-rate using the SNN given in [26]. This single layered feed-forward SNN can be trained via an exponential STDP rule to classify data-points of Fisher's Iris, Wisconsin's breast-cancer and wine data-sets. For training, following STDP rule has been used:

$$\Delta G(\Delta t, G) = \begin{cases} |A_+| \exp\left(-\frac{\Delta t}{\tau_+}\right)\left(1 - \frac{G}{G_{max}}\right)^p, & \Delta t \geq 0 \\ -|A_-| \exp\left(-\frac{\Delta t}{\tau_-}\right)\left(\frac{G}{G_{max}}\right)^p, & \Delta t < 0 \end{cases} \quad (3)$$

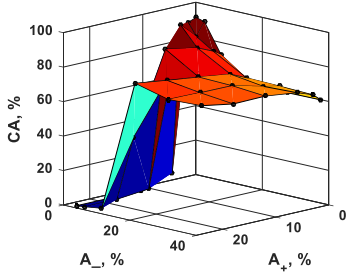

Figure 4: Classification accuracy (CA, %) versus learning-rate parameters $A_+$ and $A_-$ per cent of $w_{max}$

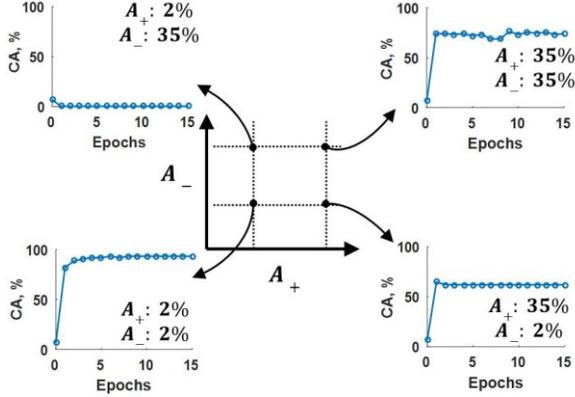

Figure 5: Classification accuracies (CA) for four values of $A_\pm$.

Where, + represents parameters for conductance potentiation, − represents parameters for conductance depression, $A_\pm$ set the maximum conductance change, $\tau_\pm$ set the STDP time-constant, $G_{max}$ is the maximum conductance (with 0 as minimum) and $p$ is conductance's saturation-factor. To quantitatively study the effect of $A_\pm$ on training, we simulate the training of the network to classify Iris dataset with various values of the pair $\{A_+, A_-\}$ and observe the evolution of classification accuracy as training proceeds.

From Figure 4, which plots the mean-accuracy for last five epochs of training, it is observed that

1. Lower the learning-rates, better the training in terms of both maximum accuracy and stability
2. Low synaptic weight's depression-rate (~2%) is a necessary condition for high (more than ~90%) post-training accuracy
3. Synaptic potentiation rate affects the training performance, but to a lower extent than depression rate

Figure 4 shows four classification accuracy (CA) plots for all possible pairs of $A_\pm \in \{2\%, 35\%\}$, where the percentages represent fraction of the maximum weight of the synapse, $w_{max}$. Following observations can be made:

1. For large $A_\pm$ (~35%), the training never gets completed and the classification accuracy doesn't settle

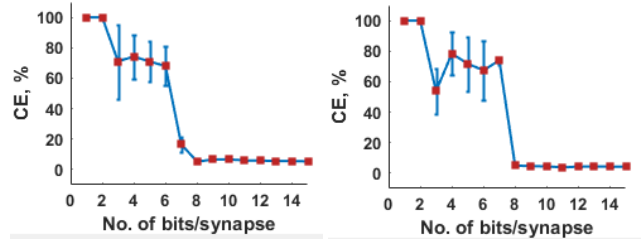

Figure 3: Distribution of classification errors (CE) versus number of conductance levels per synapse for (a) Iris dataset (b) Wisconsin's breast-cancer dataset, both trained with 4 linear sensors at input

2. For large $A_+$ but small $A_-$ (~2%), the classification accuracy settles, but is not the maximum, which means training is partial
3. For large $A_-$ but small $A_+$, the network is unable to learn
4. For small $A_\pm$, the learning is the best and the classification accuracy settles

Further, several SNNs trained using STDP in literature rely on learning-rates less than 1% [27]–[31]. Though no work focuses on how learning-rate affects the overall training, it is very likely that the performances may suffer from similar degradation as learning-rates are increased.

### C. Number of bits per digital synapses

To empirically determine the number of bits needed in a digital synapse for software-equivalent training, we simulated the training of the network used above to classify Fisher's iris and Wisconsin's breast-cancer datasets. The total conductance range was divided uniformly into $n$ levels. During a conductance-update event, the originally continuous change in conductance is discretized to the nearest conductance level. Figure 3 plots the mean post-training per-cent classification error (CE) for 10 training experiments over 10 training-epochs. It is observed that at least 8 bits/256 discrete levels are needed to ensure lowest CE.

### III. PCMO-RRAM AS SYNAPSE IN SNNS

#### A. STDP of RRAMs using neural write-pulses

RRAM can be approximately modeled as resistor whose resistance can be changed with voltage-pulses, if the pulse-strength exceeds writing threshold (Figure 6a) [32], [33]. This change increases, as strength above its threshold is increased. Given this approximation, to realize STDP on an RRAM, carefully shaped write-pulses are applied at the two ends of the synaptic RRAM by the two spiking neurons in context (Figure 6b) [19], [34]. These pulses are so shaped that when two such pulses, corresponding to each of the neurons attached to a synapse, relatively displaced in time, are subtracted,

1. There exists a portion in the net that always just reaches (or, goes above) the RRAM's threshold
2. the height of the net above the RRAM's threshold (or compactly, the overdrive) is a function of the relative displacement

Each time a neuron spikes, such pulses are applied immediately in response at the terminal of the synaptic RRAM. This way, the RRAM sees a net-voltage equal to the

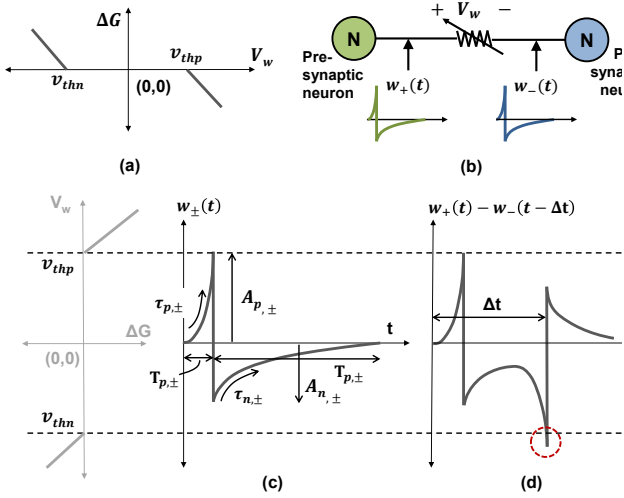

$$G_{\pm}(t) = \begin{cases} A_{p,\pm} \dfrac{(e^{t/\tau_{p,\pm}} - e^{-T_{p,\pm}/\tau_{p,\pm}})}{(1 - e^{-T_p/\tau_p})}, & -T_{p,\pm} \le t < 0 \\ A_{n,\pm} \dfrac{(e^{-t/\tau_{n,\pm}} - e^{-T_{n,\pm}/\tau_{n,\pm}})}{(1 - e^{-T_{n,\pm}/\tau_{n,\pm}})}, & 0 < t \le T_{n,\pm} \end{cases} \quad (4)$$

Here,

1. $A_p$ and $A_n$ are the positive and negative amplitudes. $A_p$ and $A_n$ are set such that they are less than $V_{tp}$ and $V_{tn}$, respectively

2. $\tau_p$ and $\tau_n$ are the decay constants. $\tau_p$ sets the write-time of the pulse and $\tau_n$ sets the exponential STDP time-constant

3. $T_p$ and $T_n$ define the spike lengths in the time axis

*B. Demonstration of exponential STDP of PCMO*

PCMO-RRAMs have been experimentally demonstrated as endurable, fast and highly scalable non-volatile analog memristive contender for synaptic applications [8], [35]–[37]. PCMO-RRAM with cross-section shown in Figure 7 and an area of $5\mu m \times 5\mu m$, originally reported in reported in [37] was used for the demonstrating STDP. The device was initialized to its low-resistance state by applying a long-lasting and constant negative voltage-pulse of –2.5 V and compliance set to 10 mA. For writing, pulses in Eq. 2 were used and the values for T's, $\tau$'s and A's are given in Table 1. The procedure followed to demonstrate the STDP of the device is:

1) the conductance was *read* using a small rectangular voltage pulse (0.5V), yielding the initial conductance value $G_0$

2) a $\Delta t$ was randomly chosen from [-100ms, 100ms] and the subtraction of relatively displaced write-pulses corresponding to $\Delta t$ was directly applied to terminal 1 and with terminal 2 grounded

3) the final conductance value ($G_f$) was recorded by using a 0.5 V voltage pulse

These three steps were repeated 1000 times for the new state $G_1$ that served as $G_0$ for the new iteration. Of the pulses applied, the ones leading to increase in conductance have been plotted in green/$\Delta$ in Figure 8and those leading to decrease, in blue/$\nabla$. The conductance values have been normalized by dividing each value with the maximum conductance observed. For a better visual, uniformly spaced iso-initial-conductance (iso-$G_i$) points have been plotted in Figure 9 and iso-time-

Figure 6: (a) RRAM conductance-change (ΔG) modeled as a function of rectangular pulse of height $V_W$, (b) Pre-synaptic and post-synaptic neuron connected through a synapse. Right after each of their spiking instants, each produce and apply a fresh write-pulse at the RRAM's terminal (c) A possible synaptic write-pulse for exponential STDP, (d) Subtraction of two relatively displace ($\Delta t > 0$) write-pulses or voltage across RRAM. The circled portion of the subtraction, crossing the threshold, changes the conductance of the RRAM

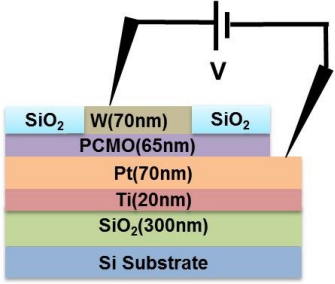

Figure 7: Measured PrCaMnO-RRAM's structure

TABLE I

WRITE-PULSE PARAMETERS USED FOR STDP

| Write-pulse | Parameter | Value | Parameter | Value |
|---|---|---|---|---|
| $w_+$ | $A_{p,+}$ | 1V | $A_{n,+}$ | -1V |
|  | $T_{p,+}$ | 1$\mu$s | $T_{n,+}$ | 100ms |
|  | $\tau_{p,+}$ | 0.5$\mu$s | $\tau_{n,+}$ | 50ms |
| $w_-$ | $A_{n,+}$ | 1V | $A_{n,-}$ | -1V |
|  | $T_{n,+}$ | 1$\mu$s | $T_{n,-}$ | 100ms |
|  | $\tau_{n,+}$ | 0.5$\mu$s | $\tau_{n,-}$ | 50ms |

subtraction of the displaced pulses applied by the pre- and the post-synaptic neurons (Figure 6d). For implementing an exponential STDP rule, the write-pulse may be given the following shape:

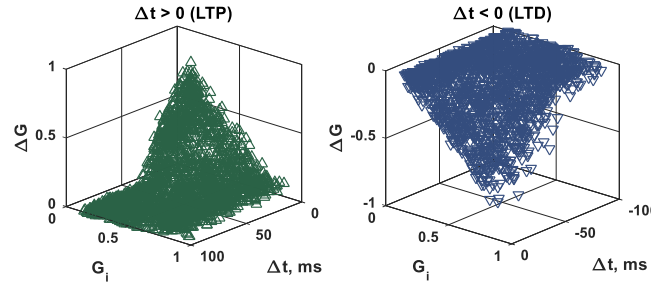

Figure 8: LTP (green/$\Delta$) and LTD (blue/$\nabla$) scatter plots

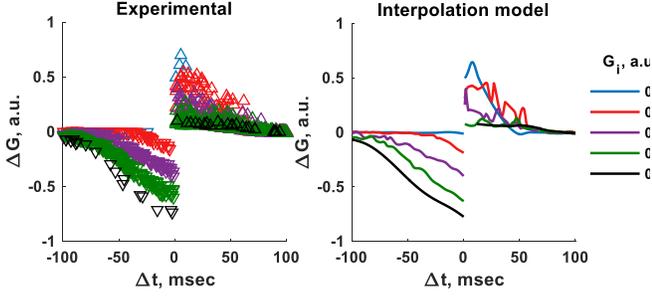

Figure 9: (a) Observed STDP (ΔG) of the PCMO-RRAM's conductance with various iso-initial-conductance ($G_i$) in distinct colors. (b) Interpolation model of measurements. $G_0$ in the legend represents the initial conductance

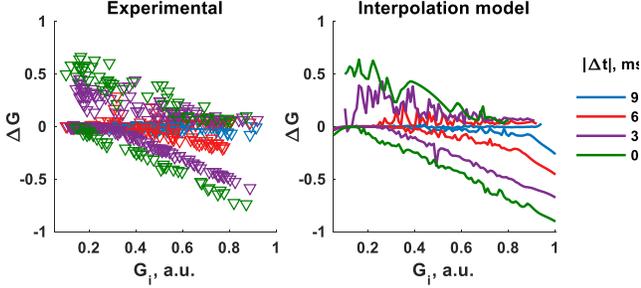

Figure 10: (a) Observed STDP (ΔG) of the PCMO-RRAM's conductance with various iso-spike-time-difference (iso-$\Delta t$) in distinct colors. (b) Interpolation model of measurements in *(a)*

difference (iso-$\Delta t$) in Figure 10. To use this data for simulations, we used its interpolation model. Iso-$G_i$ and iso-$\Delta t$ STDP curves obtained using such a model are plotted in Figure 9b and Figure 10b. Though the STDP demonstrated experimentally has a time-constant of 50ms, it can be altered by scaling the time-keeping portion of the write-pulses.

### C. Training with Single PCMO-RRAM as a synapse

We replaced the mathematical synaptic model in the network mentioned in Sec. II with the interpolation model of PCMO-RRAM. To do so, we modified the:

1. Read equation: Each read-instance of $w_{ij}$, the mathematical conductance, was replaced with $G_{RRAM}^{ij}$ the RRAM's conductance. Since the latter was normalized, an additional factor of $G_{max}$ was added, leading to the following replacement:

$$G^{ij} \rightarrow G_{max} G_{RRAM}^{ij} \quad (5)$$

2. STDP rule equation: The STDP rule specified in Eq. 3 was replaced with following equation:

$$G_{RRAM}^{ij} = G_{RRAM}^{ij} + \Delta G\left(\Delta t^{ij}, G_{RRAM}^{ij}\right) \quad (6)$$

Where, $\Delta G\left(\Delta t^{ij}, G_{RRAM}^{ij}\right)$ is determined from the interpolation model mentioned above.

Next, the network's training was simulated. Figure 10 plots the evolution of classification accuracy as training proceeds. Clearly, the performance is worse in comparison to that obtained with a mathematical synaptic model.

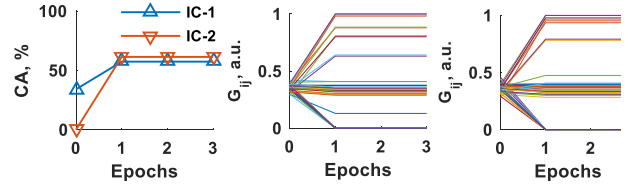

Figure 11: Training behaviour with single RRAM based synapse. (a) Regardless of the initial condition post-training classification accuracies (CA) settle to low values right after first training iteration (epoch); (b) Conductance's evolution with initial condition (IC)-1; (c) Conductance's evolution with IC-2.

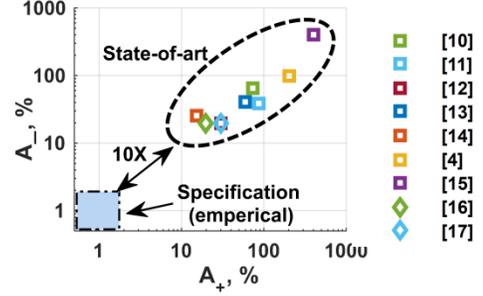

Figure 12: Learning-rate ($A_+$ or $A_-$) offered by various analog RRAMs in literature. None of them meet our specifications

Our hypothesis, based on observation made from Figure 4 and Figure 11, is that such marked reduction in performance is observed because of large percentage change in conductance of PCMO-RRAM. This is equivalent to having large learning-rates in a mathematical model.

STDP has been demonstrated on several analog RRAMs [4], [10]–[17]. As observed above, it is necessary for the maximum conductance change to be less than 2% of the synaptic conductance-range to get software-equivalent evolution of conductance. However, all analog conductance's plasticity demonstrations up until now have a maximum conductance change ($\max_{\Delta t}(|\Delta G|/G)$) of more than 20% and can go up to 400% in some devices (Figure 12). Thus, all currently existing analog RRAMs will fail to produce software-equivalent post-training classification accuracy for our network.

Our second hypothesis, based on observation made from Figure 6 and 7, is that training performance can improved without changing the network, by *reducing* the maximum change in conductance change (i.e., lower $|\Delta G(0^+, G_{min})|$ or $|\Delta G(0^-, G_{max})|$ in Eq. 3). The validity of this hypothesis can be tested by using an STDP-data like we used for an RRAM with a lower maximum conductance. Since in our knowledge, all RRAMs in the literature do not meet this constraint, this is not possible for us at the moment. To test our hypothesis and as a step towards better memristor based synaptic models, we propose using a set of n PCMO-RRAMs in parallel in a way that

1. they function aggregately as synapse
2. the conductance-change pulses are applied to only one of n-RRAMs

Next, we validate our proposal.

## IV. PARALLEL AND MULTIPLE PCMOS AS SYNAPSE

To test the ability of multiple (n) PCMO-RRAMs to collectively act as a synapse, we continued with same network and trained it to classify Iris data-set. The mapping from a mathematical synapse to an n-RRAM based synapse is similar to the one described in Sec. 3C, with a slight difference, described as follows:

1. Each read-instance of $G_{ij}$, the mathematical weight, was replaced with:

$$G^{ij} \to \frac{G_{max}}{M} \sum_{k=1}^{M} G_{RRAM}^{k,ij} \quad (7)$$

2. STDP rule equation: For every spiking instant, STDP-rule based weight-update is replaced with the following two equations:

$$K = rand(1,2\ldots M)$$
$$G_{RRAM}^{K,ij} = G_{RRAM}^{K,ij} + \Delta G(\Delta t^{ij}, G_{RRAM}^{K,ij}) \quad (8)$$

Note that RRAM's conductance increase corresponds to *increase* in the conductance of the synapse of which it's a part. We let $n$ take all values from the set {2, 4, 16, 36, 64, 100}. Figure 13 plots the CA, as training proceeds for n=2, 16 and 100. It is observed that learning is more stable for synapses with more RRAMs. Figure 13 plots the CA's quantiles for all $n$'s, as the training progresses with number of training-epochs in Figure 14. The number of training-epochs was determined empirically. For n=100, approximately 50 epochs were needed to stabilize learning. Since learning-rate is roughly inversely proportional to the number of RRAM, the number of training iterations needed is proportionally increased/decreased. For very large learning-rates (n=2, 4 and 16), a lower limit of 20 was set. The evolution conductance of each synapse for various $n$'s has been plotted in Figure 15. It is observed that:

1. starting from n=4, the network's CA reaches software-maximum of 97.3% at least *once*, while training (Figure 13)
2. despite being trained for more than adequate epochs, network with synapses having very low number of RRAMs are unable learn *stably* (Figure 13 and Figure 14a)
3. as n is increased, the CA distribution in Figure 14a follows a trend similar to the one exhibited by Figure 5 showing the increase in CA as LTP and LTD rates are decreased simultaneously – higher mean and lower variation
4. conductance evolve more smoothly as $n$ increases

## V. DISCUSSION

The scheme discussed does not escalate the circuit requirements. During reading, all RRAMs in a synapse need to be simultaneously read from. This is done by applying same reading pulse-voltage to all row-bars of the synapse. The current from each of the branch associated to a synapse is then summed up, to get a current proportional to the synaptic weight.

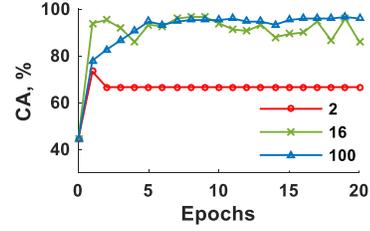

Figure 13: Classification accuracies for 2, 16 and 100 PCMO-RRAMs in the synapse, with same initial conductance configuration

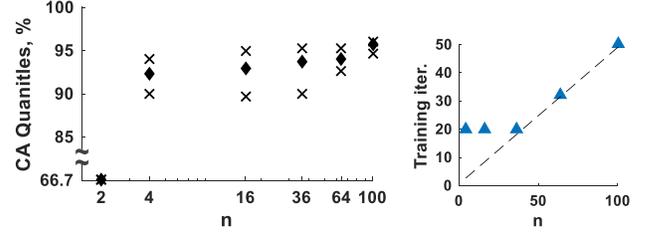

Figure 14: (a) Classification-accuracy (CA) quantiles as the number of PCMO-RRAMs in the synapse is varied. (b) Number of epochs for training

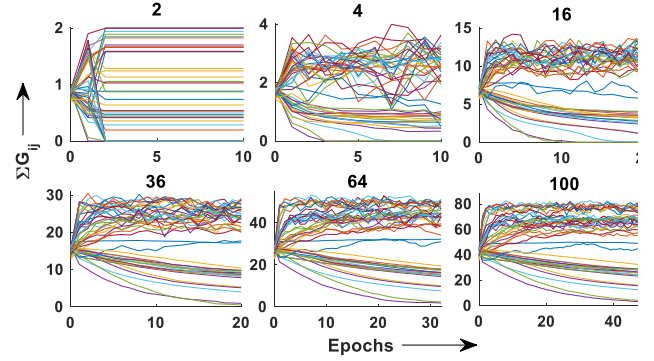

Figure 15: Synaptic weight evolution smoothens as more PCMO-RRAMs are added within the synapse

For performing the writing operation, one RRAM is randomly chosen from $M$ RRAMs within a synapse, with a uniform probability for each. This is done by applying the pre-synaptic write-pulse to a random row among $M_1$ row-bars and the post-synaptic write-pulse to a random column among $M_2$ column-bars (for a synapse with $M_1$ RRAM rows and $M_2$ RRAM columns). This way, the RRAM selection is uniformly random among all $M$ RRAMs. Though there can be other schemes for uniformly selecting RRAMs, this in our opinion, shouldn't require complicated and/or large area peripheral circuits.

The reading and writing phases may be multiplexed in time using a global control signal $RW$ (reading indicated by $RW = 1$ and writing by $RW = 0$). During the writing phase, to allow random selection of a row/column, a set of global one-hot selection-lines is laid out along the periphery of the array. The active line in this set is changed, periodically. Whenever a neuron ($N_i$) spikes (assumed to be random in time), the content of the selection line set is copied onto the RRAM selection register with output vector $\bar{S}_i$ (Figure 17). This way, the same row/column of the synapse remains selected until the next spike, as the active line of the synapse's selection- register remains same. During the reading phase, the selection register is over-written using the pre-set control input so that $\overline{S_i} = \{1,1 \dots 1\}$.

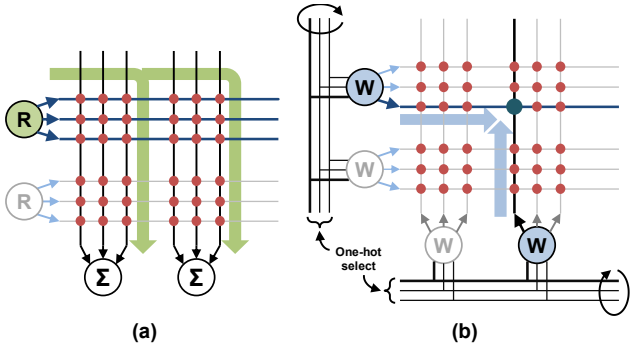

Figure 16: (a) Read process consists of applying read pulses (R, in green) to all rows of a synapse and summing up current at the array's virtually grounded output; (b) Weight update process consists of application of write-pulses (generated by W, blue) to only one randomly chosen row and column of the synapse. Choice of the row/column is determined by global one-hot select lines that are read at the spiking instant

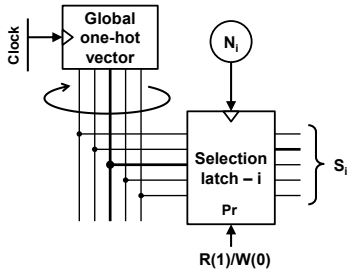

Figure 17: At every rising edge of neural spike, selection-latch copies content of the global one-hot vector. For each neuron, there is a unique selection latch

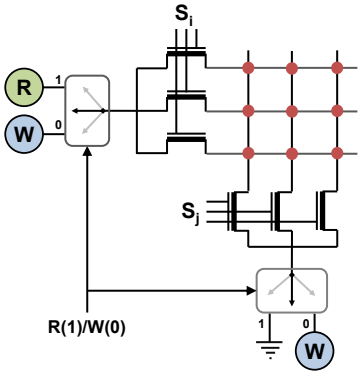

Figure 18: During writing phase (RW=0), $n$ MOSFET-based switches select the RRAM being written onto. During reading phase (RW=1), all RRAMs are accessed by turning on all switches.

Write-pulses may be applied to a single row (column) out of M1 (M2) via MOSFET based switches as shown in Figure 18. Each of the MOSFETs' gate is connected to the one-hot selection vector $S_i$. Thus, only one MOSFET out of M1 (M2) is conducting and will allow write-pulses to be applied to the corresponding row (column). However, during the reading phase ($RW = 1$), all MOSFETs are turned on by setting $\overline{S_i} = \{1,1 \dots 1\}$. Thus, for all RRAMs within a synapse, same reading pulse is applied on pre-synaptic terminals and post-synaptic terminals grounded.

Use of multiple RRAMs within a synapse clearly requires a bigger cross-bar array. However, a larger cross-bar implies a larger attenuation in the voltage applied across RRAMs at cross-points far away from the either input and output sides of the array. Thus, for a constant wire-resistance, the cross-bar array size, or more fundamentally, the number of RRAMs in the synapses cannot be made arbitrarily large. Not only their number, but also their arrangement needs to be carefully designed to maintain a certain minimum level of read-write fidelity. For simplicity, consider an SNN with just one n-RRAM synapse and all RRAMs within it in same conductance state. If the potential-drop due to wire-resistance is assumed to be a linear function of cross-point index within a synapse, then the read current-error of a cross-point $(r, c)$ within the array can be expressed as:

$$\delta I(r,c) \propto r + c \qquad (9)$$

For a synapse with $R \times C$ RRAMs, the maximum error in current will be for the cross-point at the corner furthest from the inputs and outputs, i.e., the one with indices $(R, C)$. For fixed number of RRAMs in a synapse, say $RC$, the maximum error in read current will be

$$\delta I_{max} = \max_{RC=n}(R + C) \qquad (10)$$

This happens when the synapse has an arrangement of $RC$ cross-points in configuration closest to that of a square. Thus, within a synapse, the RRAMs should be arranged in a square-like configuration.

## VI. CONCLUSION

In this work, we introduce and empirically determine two new specifications for an SNN based synapse: number of conductance levels per synapse and maximum learning-rate. To the best of our knowledge, there are no RRAMs that meet the latter specification. As a solution, we proposed the use of multiple RRAMs in parallel within a synapse. While synaptic reading, all RRAMs are simultaneously read and for each synaptic conductance-change event, the writing pulses for STDP are applied on only one RRAM, randomly picked from the set. To validate our solution, we experimentally demonstrated STDP of conductance of a PCMO-RRAM and showed that due to large learning-rate, a single device fails to model a synapse in the training of an SNN. As anticipated, network training improved as more RRAMs are added to the synapse. Lastly, we discuss the circuit-requirements for implementing such a scheme, to conclude that the requirements are within bounds.